\documentclass[fleqn,usenatbib]{mnras}

\usepackage{newtxtext,newtxmath}

\usepackage[T1]{fontenc}

\DeclareRobustCommand{\VAN}[3]{#2}
\let\VANthebibliography\thebibliography
\def\thebibliography{\DeclareRobustCommand{\VAN}[3]{##3}\VANthebibliography}

\usepackage{graphicx}	
\usepackage{amsmath}	

\usepackage[usenames,dvipsnames]{xcolor}
\usepackage{xspace}

\newcommand{\mas}{\textit{mas}\xspace}
\newcommand{\Teff}{T_\text{eff}}
\newcommand{\logL}{\log(L/\mathrm{L}_\odot)}
\newcommand{\Msunyr}{\mathrm{M}_\odot\,\text{yr}^{-1}}
\newcommand{\kms}{\text{km}\,\text{s}^{-1}}

\defcitealias{Kilpatrick23}{K23}
\defcitealias{Soraisam23}{S23}
\defcitealias{Jencson23}{J23}

\title[SN 2023ixf Progenitor]{The Progenitor Star of SN 2023ixf:
A Massive Red Supergiant with Enhanced, Episodic Pre-Supernova Mass Loss}

\author[Y.-J. Qin et al.]{
Yu-Jing Qin,$^{1}$\thanks{E-mail: yujingq@caltech.edu}
Keming Zhang,$^{2,3}$
Joshua Bloom,$^{2}$
Jesper Sollerman,$^{4}$
Erez A. Zimmerman,$^{6}$ \newauthor
Ido Irani,$^{6}$
Steve Schulze,$^{4}$ 
Avishay Gal-Yam,$^{6}$
Mansi Kasliwal,$^{1}$
Michael W. Coughlin,$^{7}$ \newauthor
Daniel A. Perley, $^{8}$
Christoffer Fremling,$^{1,9}$
Shrinivas Kulkarni$^{1}$
\\
$^{1}$Division of Physics, Mathematics and Astronomy, California Institute of Technology, 1200 E California Blvd., Pasadena, CA 91125, USA \\
$^{2}$Department of Astronomy, University of California, Berkeley, 501 Campbell Hall \#3411, Berkeley, CA 94720-3411, USA \\
$^{3}$Department of Astronomy and Astrophysics, University of California, San Diego, 9500 Gilman Dr., La Jolla, CA 92093, USA \\
$^{4}$Department of Astronomy, The Oskar Klein Center, Stockholm University, AlbaNova University Center, SE 106 91 Stockholm, Sweden \\
$^{5}$Department of Physics, The Oskar Klein Centre, Stockholm University, Albanova University Center, SE 106 91 Stockholm, Sweden \\
$^{6}$Department of Particle Physics and Astrophysics, Weizmann Institute of Science, 234 Herzl Street, POB 26, Rehovot 7610001, Israel \\
$^{7}$School of Physics and Astronomy, University of Minnesota, Twin Cities, 116 Church Street S.E., Minneapolis, MN 55455, USA \\
$^{8}$Astrophysics Research Institute, Liverpool John Moores University, Liverpool Science Park, 146 Brownlow Hill, Liverpool L3 5RF, UK \\
$^{9}$Caltech Optical Observatories, California Institute of Technology, 1200 E California Blvd., Pasadena, CA 91125, USA\\
}

\date{Accepted XXX. Received YYY; in original form ZZZ}

\pubyear{2023}

\begin{document}
\label{firstpage}
\pagerange{\pageref{firstpage}--\pageref{lastpage}}
\maketitle

\begin{abstract}
We identify the progenitor star of SN~2023ixf in the nearby galaxy Messier 101 using Keck/NIRC2 adaptive optics imaging and pre-explosion {\it HST}/ACS images.
The supernova position, localized with diffraction-spike pattern and high precision relative astrometry, unambiguously coincides with a single progenitor candidate of $m_\text{F814W}=24.96_{-0.04}^{+0.05}$.
Forced photometry further recovers $2\sigma$ detections in the F673N and F675W bands and imposes robust flux limits on the bluer bands.
Given the reported infrared excess and semi-regular variability of the progenitor, we fit a time-dependent spectral energy distribution (SED) model of a dusty red supergiant (RSG) to a combined dataset of {\it HST} photometry, as well as ground-based near-infrared and {\it Spitzer}/IRAC [3.6], [4.5] photometry from the literature.
The progenitor closely resembles a RSG of $T_\text{eff}=3343\pm27$ K and $\logL=5.10\pm0.02$, with a $0.11\pm0.01$ dex ($25.2\pm1.7$ per cent) variation over the mean luminosity at a period of $P=1128.3\pm6.5$ days, heavily obscured by a dust envelope with an optical depth of $\tau=2.83\pm0.03$ at $1\,\mu\text{m}$ (or $A_\text{V}=10.28\pm0.11$ mag).
Such observed signatures match a post-main sequence star of $18.1_{-1.2}^{+0.7}\,\mathrm{M}_\odot$, close to the most massive SN II progenitor, with a pulsation-enhanced mass-loss rate of $\dot{M}=(3.58\pm0.15)\times 10^{-4} \,\Msunyr$.
The dense and confined circumstellar material is likely ejected during the last episode of radial pulsation before the explosion.
Notably, we find strong evidence for periodic variation of $\tau$ (or both $\Teff$ and $\tau$) along with luminosity, a necessary assumption to reproduce the wavelength dependence of the variability, which implies dust sublimation and condensation during radial pulsations.
Given the observed SED, partial dust obscuration remains a possible scenario, but any unobstructed binary companion over $7.1\,M_\odot$ can be ruled out.
\end{abstract}

\begin{keywords}
transients: supernovae -- stars: supergiants -- supernovae: individual: SN~2023ixf 
\end{keywords}

\section{Introduction}

\begin{figure*}
\includegraphics[width=\textwidth]{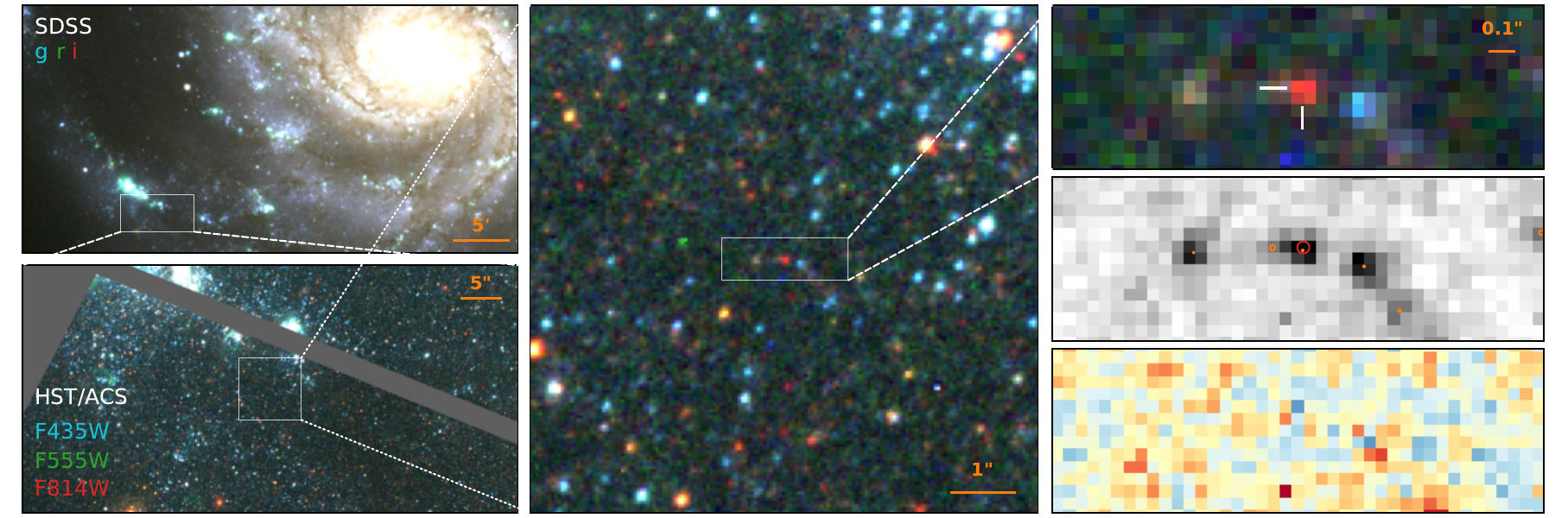}
\caption{The progenitor star of SN~2023ixf in its host galaxy, Messier 101. The upper left panel shows a Sloan Digital Sky Survey (SDSS) cutout, and the following zoom-in panels show the {\it HST}/ACS colour composite image near the SN.
The upper right panel indicates the SN position localized with Keck/NIRC2 adaptive optics image. The middle panel shows the sources detected in the pre-explosion {\it HST}/ACS image with yellow circles showing the $1\sigma$ error of source positions, and the red circle indicating our Keck/NIRC2 localization error (including systematic error).
Finally, the lower right panel shows the residual map ($-3$ to $3$ times the background RMS) after the source detection procedure.
The SN position localized with our Keck/NIRC2 image unambiguously coincides with the red source indicated by the crosshair in the upper right panel.}
\label{fig:fig1}
\end{figure*}

Connecting the diverse supernova (SN) phenomena to the properties and late-stage evolution of progenitor stars is a pivotal task in the study of stellar transients. Deep and high-resolution pre-explosion images of nearby host galaxies, particularly acquired with the Hubble Space Telescope ({\it HST}) over its three decades of operation, remains the only \textit{direct} approach to constrain SN progenitor properties.

Currently, there are about $30$ core-collapse supernovae (CCSNe) with direct progenitor detections \citep[e.g.,][]{Smartt15, VanDyk17} and the majority of them are Type II supernovae (SNe II), the most abundant SN subclass by volumetric rates \citep[e.g.,][]{Li11, Shivvers17}.
SN II progenitors retain part of their hydrogen-rich envelopes before the explosion, giving rise to Balmer emission lines in their photospheric-phase spectra. Direct progenitor detections broadly support the scenario that single massive stars with Zero-Age Main Sequence mass (or simply ``mass'') of about $8$ to $18\,\mathrm{M}_\odot$ explode as SNe II during the Red Supergiant (RSG) phase (e.g., \citealt{Smartt09, Smartt15}) -- the only SN subclass that has been reliably connected to a specific type of progenitor star so far.

Despite the success, the RSG-to-SN II connection is not yet completely understood.
A major unsettled issue is that directly detected SN II progenitors are rarely more luminous than $\log (L/\mathrm{L}_\odot)\sim 5.1$ (or more massive than $\sim 18\,\mathrm{M}_\odot$), but the observed RSG population in the Milky Way and nearby galaxies extends to $\log (L/\mathrm{L}_\odot) \sim 5.5$ (or $\sim 25$ -- $30\,\mathrm{M}_\odot$), a discrepancy commonly referred to as the ``RSG problem'' (\citealt{Smartt09, Smartt15}; but see also \citealt{Davies18, Davies20}).
The absence of SN II progenitors above $\logL\sim5.1$ could be attributed to the direct collapse of massive RSGs into black holes (e.g., \citealt{OConnor11, Horiuchi14}), the increased circumstellar extinction near massive RSGs (\citealt{Walmswell12, Beasor16}), or a ``superwind'' phase that removes the hydrogen-rich envelopes and produces stripped-envelope supernovae instead of SNe II (\citealt{Yoon10}; but see also \citealt{Beasor22}).
Expanding the existing sample of directly detected SN II progenitors, determining their luminosity, mass, and mass-loss rate, and identifying limiting cases of progenitor properties would be vital to test these possible scenarios.

The nearby SN~2023ixf in the galaxy Messier 101 provides a once-in-a-decade opportunity to take a closer look at a SN II progenitor through the rich pre-explosion data.
SN~2023ixf was reported by \citet{Itagaki23TNS} on 2023 May 19 at 21:42 UT and was spectroscopically classified by \cite{Perley23TNS}.
Early follow-up campaign focuses on the dense and confined circumstellar material (CSM) probed by flash ionizing features in the optical \citep{Bostroem23, Hiramatsu23, JacobsonGalan23, Smith23, Vasylyev23, Yamanaka23} and ultraviolet \citep{Teja23, Zimmerman23} wavelengths, as well as neutral hydrogen absorption in hard X-ray \citep{Grefenstette23}. Remarkably, both spectropolarimetry \citep{Vasylyev23} and high-resolution spectroscopy \citep{Smith23} suggest asymmetric CSM distribution.
Meanwhile, photometric analyses \citep{Hiramatsu23, Hosseinzadeh23, JacobsonGalan23, Teja23, Zimmerman23} also require CSM interaction as an additional power source of SN emission.
It has been suggested that the dense and confined CSM leads to an extended shock breakout phase \citep{Hiramatsu23, Hosseinzadeh23, Zimmerman23}.
Being a bright nearby SN, small telescope arrays \citep{Bianciardi23, Sgro23} and amateur astronomers \citep{Mao23TNS, Yaron23TNS} have contributed valuable early-time data of SN~2023ixf.
Furthermore, the absence of submillimeter \citep{Berger23}, X-ray \citep{Panjkov23}, Gamma-ray \citep{Sarmah23, Muller23}, and neutrino \citep{Sarmah23, Guetta23} detections at early time also impose constraints on the progenitor properties and the detailed physical processes of CCSNe.

Finally, the physical properties of a candidate progenitor star have been discussed in several recent papers \citep{Jencson23, Kilpatrick23, Niu23, Pledger23, Soraisam23, VanDyk23_23ixf}.
The red optical colour, strong infrared excess, and semi-regular variability of the candidate indicate a luminous RSG with radial pulsations, heavily obscured by circumstellar dust.
Retrospective analysis of archival optical \citep{Dong23, Neustadt23, Panjkov23}, ultraviolet \citep{Flinner23, Panjkov23}, and X-ray \citep{Panjkov23} data rules out bright outbursts and eruptive mass loss of the candidate.

In this work, we localize the progenitor star of SN~2023ixf using high-resolution adaptive optics imaging.
We also constrain its progenitor properties by fitting pre-explosion photometry with the SED of a dusty RSG.
We confirm and strengthen the identification of the progenitor in previous works and demonstrate that the progenitor is close to the empirical luminosity upper limit of SN II progenitors.
Through the paper, we use a distance to the host of $D=6.90\pm 0.12$ Mpc (or $\mu=29.194\pm0.039$ in distance modulus; \citealt{Riess22}), and luminosity values are calculated with distance uncertainty folded in. Source brightness, if in magnitude scale, is reported in the AB magnitude system.

\begin{figure*}
\includegraphics[width=\textwidth]{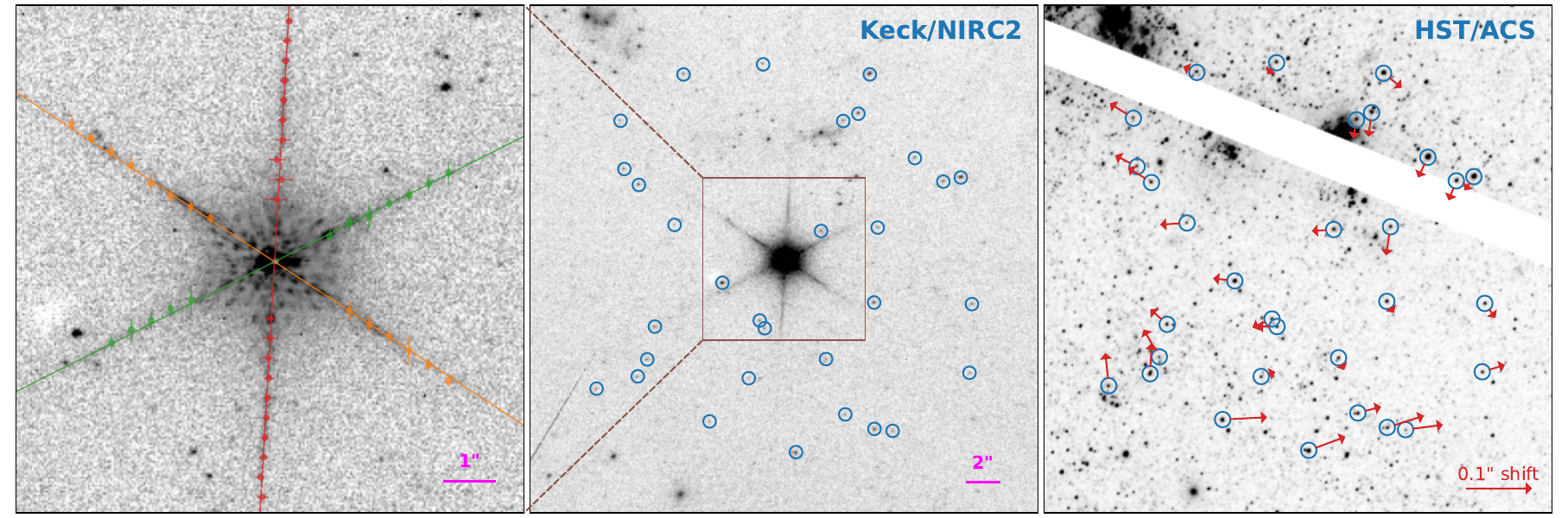}
\caption{Localizing SN~2023ixf in the pre-explosion {\it HST}/ACS image using Keck/NIRC2 adaptive optics imaging.
The left panel shows a contrast-enhanced cutout of the post-explosion Keck/NIRC2 Ks-band image (middle panel, in north-up, east-left orientation), centred at the heavily saturated SN.
The SN location is precisely determined by fitting the positions of diffraction spikes (colour points with error bars) using a simple linear pattern (colour lines).
The middle panel shows the astrometric reference stars (blue circles) detected in the Keck/NIRC2 image, while the right panel shows the same field of the three-band combined {\it HST}/ACS image.
Fitting an astrometric solution with these stars, the SN can be localized with a total uncertainty of $19.5$~\mas in the {\it HST}/ACS image.
Red arrows indicate the residual error of the best-fitting astrometric solution, which dominates the error budget.
}
\label{fig:fig2}
\end{figure*}

\section{Data}

\subsection{Pre-explosion {\it HST} observations}

The SN location has been imaged by several {\it HST} programs before the explosion, including Proposal IDs 6829 (PI: You-Hua Chu), 9490 (PI: Kip Kuntz), 9720 (PI: Rupali Chandar), 13361 (PI: William Blair), and 15192 (PI: Benjamin Shappee), with a variety of instrument and band combinations, covering a time frame of nearly three decades.

We access the calibrated science images from these programs at the Mikulski Archive for Space Telescopes.\footnote{\url{https://archive.stsci.edu/}}
The data archive also provides single-visit mosaics for the Wide Field Camera~3 (WFC3) and Advanced Camera for Surveys (ACS) programs. We choose the stacked image from Proposal ID 9490, which combines 2340 seconds of exposure in the F435W, F555W, and F814W bands and are aligned with Gaia sources \citep{Gaia18}, as our detection image and astrometric reference frame.
Due to the limited pointing repeatability of {\it HST}, we also register calibrated science images to this reference image using \texttt{TweakReg} in \textsc{drizzlepac}\footnote{\url{https://github.com/spacetelescope/drizzlepac}} so they share the same astrometric reference frame and can be used for forced PSF photometry later.

\subsection{Adaptive Optics imaging with Keck/NIRC2}

We imaged the field of SN~2023ixf on 2023 May 25 at 11:16 UT (6.6 days after the explosion) using the Near-Infrared Camera (NIRC2) with Natural-Guide-Star (NGS) Adaptive Optics on the W. M. Keck II telescope, under Program ID U152 (PIs: Bloom, Zhang). To increase the overlap with the pre-explosion {\it HST}/ACS image and hence the number of usable astrometric reference stars, we choose the wide camera mode ($40$ arcsec square field, $0.0397\,\text{arcsec}\,\text{pix}^{-1}$). We acquired three 60-second science images in the K-short ({\it Ks}) band and an additional 60-second image at a nearby empty field for sky background and dark current subtraction. The science images are then sky-subtracted, flat-corrected, and averaged into a single image.
We also create a per-pixel uncertainty map with \texttt{calc\_total\_error} implemented in \textsc{photutils} \citep{Bradley22}, using the instrument gain.
The observing setup here allows us to detect fainter astrometric reference stars, but the SN itself becomes inevitably saturated due to the dramatic contrast in brightness between the SN and other stars in the field. 
We localize the SN with diffraction spikes, as described in the next section.

\subsection{Infrared photometry from the literature}

Messier 101, the host galaxy of SN~2023ixf, has been continuously monitored by the {\it Spitzer} Space Telescope over the past two decades.
Retrospective analysis of {\it Spitzer}/IRAC data at the SN position revealed the semi-regular variability of a likely progenitor in the [3.6] and [4.5] bands (\citealt{Jencson23, Kilpatrick23, Soraisam23}; hereafter \citetalias{Jencson23, Kilpatrick23, Soraisam23}), with an amplitude of $70$ per cent and a period of $1194$ days \citepalias{Jencson23}.
The {\it Spitzer} source is coincident with the best-localized SN position based on our high-resolution Keck/NIRC2 image.
To constrain the stellar and CSM properties of the progenitor, we obtain the {\it Spitzer}/IRAC measurements in \citetalias{Jencson23}. The reported Vega magnitudes are converted to flux densities (in $\mu\text{Jy}$) based on the zero-magnitude fluxes in the IRAC Instrument Handbook.\footnote{\url{https://irsa.ipac.caltech.edu/data/SPITZER/docs/irac/}}
Since the source is undetected in the {\it Spitzer}/IRAC [5.8] and [8.0] bands, we do not include measurements in these bands.

The source is also detected in archival ground-based near-infrared (NIR) images \citepalias{Jencson23, Kilpatrick23, Soraisam23}, which reveal similar variability \citepalias{Jencson23, Soraisam23} with a potentially greater amplitude than in the {\it Spitzer}/IRAC bands \citepalias{Soraisam23}.
For our analysis, we compiled NIR magnitudes from several sources:
1) one epoch of \textit{J} and \textit{K}-band magnitudes in \citetalias{Kilpatrick23}, based on the Gemini Near-Infrared Imager (NIRI) data; 2) one epoch of \textit{Ks}-band magnitude in \citetalias{Kilpatrick23}, based on the Mayall 4-meter Telescope NOAO Extremely Wide Field Infrared Imager (NEWFIRM) data; 3) eight epochs of \textit{J}, \textit{H}, and \textit{K}-band magnitudes in \citetalias{Soraisam23}, based on the archival data of UKIRT Wide Field Camera (WFCAM); and 4) five epochs of \textit{J} and \textit{Ks}-band magnitudes in \citetalias{Jencson23}, based on the MMT and Magellan Infrared Spectrograph (MMIRS) data.
The reported magnitudes are also converted to flux densities.
We choose the zero magnitude flux densities of the Two Micron All Sky Survey (2MASS; \citealt{Skrutskie06}) since the reported Vega magnitudes are calibrated with 2MASS sources.

\begin{figure*}
\includegraphics[width=0.5470085470085471\linewidth]{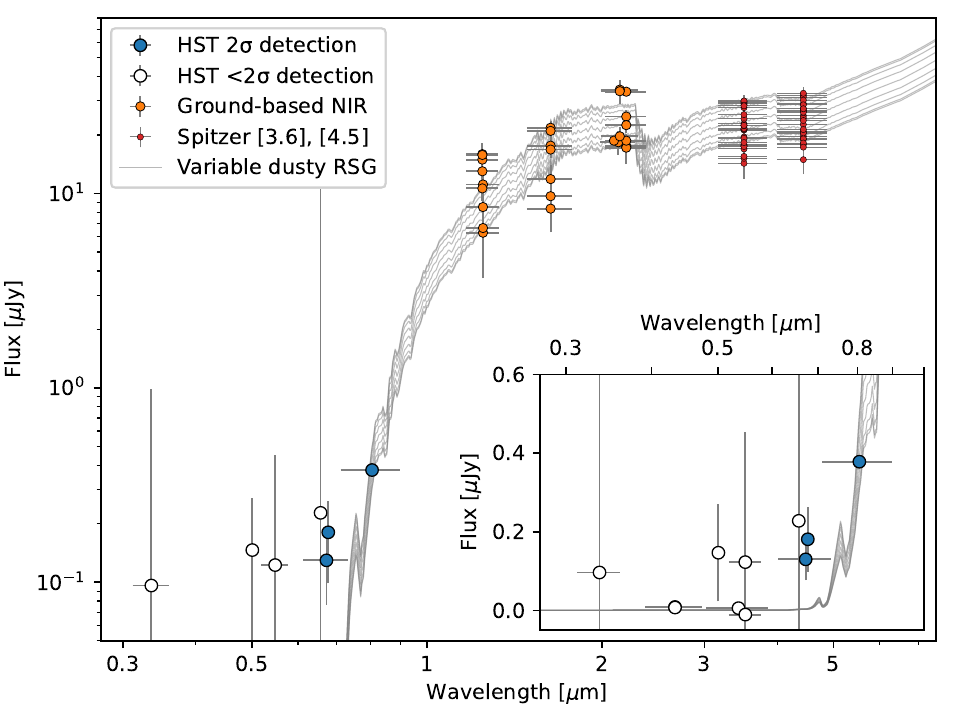}
\includegraphics[width=0.1367521367521368\linewidth]{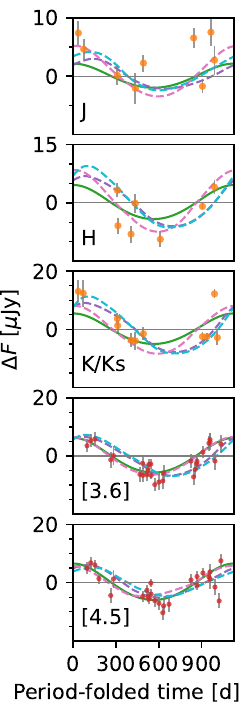}
\includegraphics[width=0.3076923076923077\linewidth]{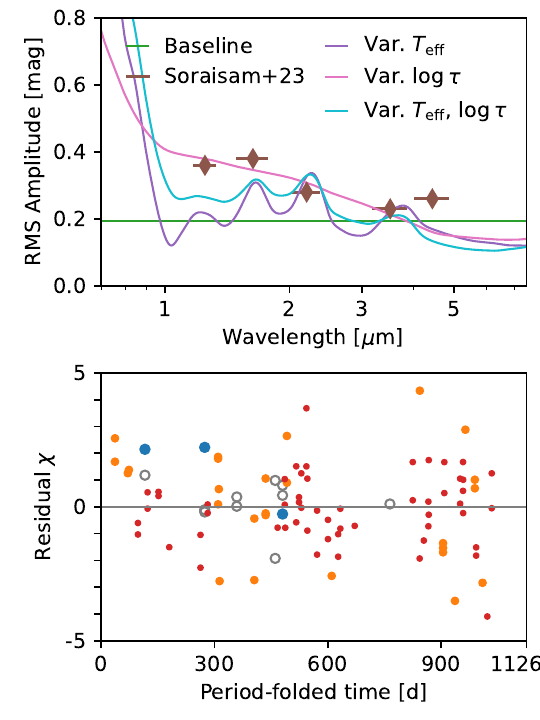}
\caption{
Modelling the observed SED with a variable dusty RSG model.
The left panel shows {\it HST} optical and literature-compiled infrared photometry (data points) and the best-fitting baseline model (grey lines) assuming a variable $\logL$ but constant $\Teff$ and $\log \tau$. The inset panel shows the optical part on a linear scale.
The middle panel shows the mean-subtracted and period-folded flux densities (data points) and the predictions of our baseline and alternative models (curves, the same colours as in the upper-right panel).
The upper right panel shows the predicted amplitude--wavelength relationship of the baseline and alternative models (Section \ref{sec:properties}), compared to the amplitude measured in \citetalias{Soraisam23}.
Alternative models with variation of $\log \tau$ (or $\Teff$ and $\log \tau$) better reproduce the observed increase of amplitude towards shorter wavelengths.
Finally, the lower right panel shows the residual error (model-observation difference, normalized by errors) for the baseline model fit (left panel), as a function of period-folded time.
}
\label{fig:fig3}
\end{figure*}

\section{Analysis and Results}

\subsection{Astrometric localization}

To identify the progenitor star in the pre-explosion {\it HST}/ACS image, we first locate the exact pixel position of the SN in the Keck/NIRC2 image, and then transform the pixel position to the {\it HST}/ACS image.

Given that the SN is saturated, we localize its pixel position using the diffraction pattern caused by the hexagonal mirror segments (Figure \ref{fig:fig2}, left). Since the primary is symmetric, the instrument is on-axis, and the field is centred at the SN, the spikes should intersect at the SN location.
To determine the X-axis positions of the north-south spike and Y-axis positions of the northeast-southwest and northwest-southeast spikes, we extract one-dimensional (1-d) light profiles and the associated uncertainty along adjacent horizontal or vertical slices with a width of $10$ pix, then fit the 1-d light profiles with a Gaussian component on a linear background.
The peak of the Gaussian component and the uncertainty represent the spike position along the slice. Erroneously determined spike positions, due to either the absence of a peak or the existence of other sources, are removed during visual inspection.
We fit the remaining 54 spike positions using three lines separated by $60\degr$ sharing a common intersection point. The free parameters are the central position ($x_\text{c}$, $y_\text{c}$) and the position angle ($\theta$) of the entire pattern.
We use \textsc{emcee} \citep{ForemanMackey13}, a Markov chain Monte Carlo (MCMC) sampler, to carry out the fit. Upon convergence, the SN position in the Keck/NIRC2 image is determined down to a statistical error (geometric mean of uncertainties in $x_\text{c}$ and $y_\text{c}$) of $0.04$ pix or $1.6$ milliarcseconds (\mas).

We then fit a relative astrometric solution across Keck/NIRC2 and {\it HST}/ACS images to transform the SN position back onto the pre-explosion image.
First, we detect point sources above a signal-to-noise ratio (SNR) threshold of $5$ in the Keck/NIRC2 image and $10$ in the {\it HST}/ACS image, with the \texttt{DAOFind} algorithm implemented in \textsc{photutils} \citep{Bradley22}.
Within $15$ arcsec from the SN position, we choose $31$ unambiguous and isolated point sources in the Keck/NIRC2 image that are also detected in the {\it HST}/ACS image as reference stars.
The astrometric solution is obtained by fine-tuning the central RA/Dec, orientation, and pixel scale of the Keck/NIRC2 image so the predicted pixel position of reference stars, based on the measured sky coordinates in the {\it HST}/ACS image and the fine-tuned World Coordinate System (WCS) parameters, best matches their detected positions.
The transformation parameters are estimated using \textsc{emcee}, where the inverse variance-weighted sum of squared residual distances is minimized.
Upon convergence, the central coordinate of the Keck/NIRC2 image is determined down to an uncertainty of $5.6$~\mas, which we consider the statistical error of the astrometric solution.
For the eight nearest reference stars within $7.3$ arcsec to the SN position, the mean residual error of the astrometric solution is $18.6$~\mas, which we consider the systematic error of the astrometric solution.
As a cross-check, we also fit an Affine transformation of pixel positions across Keck/NIRC2 and {\it HST}/ACS images including translation, rotation, and scaling, with the same set of reference stars. We obtain consistent central coordinates within the statistical errors.

The SN position localized by fitting diffraction spikes, after transformed to the pre-explosion color-composite {\it HST}/ACS image, points to a red source (Figure \ref{fig:fig1}) with a total uncertainty of $19.5$~\mas.
The red source is also the most likely progenitor candidate proposed in earlier works, including the localization by \citet{VanDyk23_23ixf} using 'Alopeke imaging.

\begin{figure*}
\includegraphics[width=0.4571428571428571\linewidth]{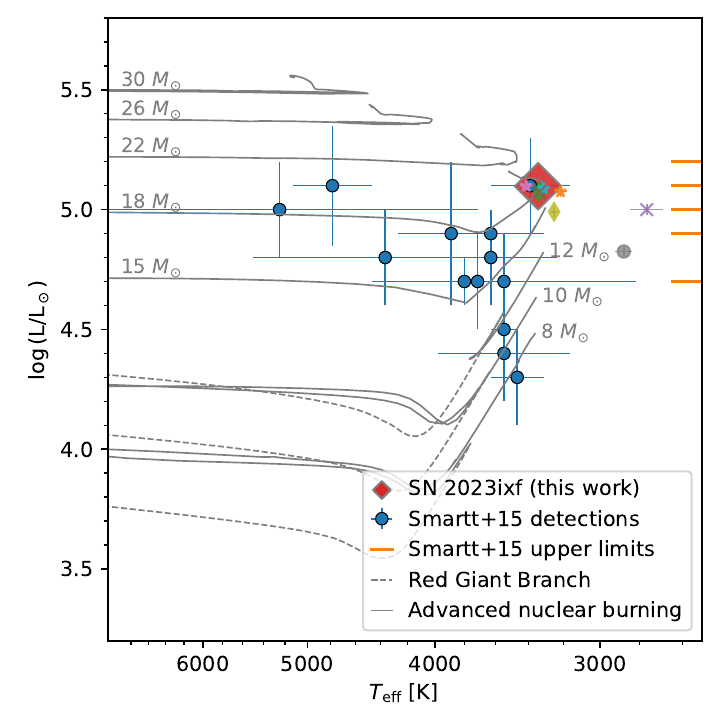}
\includegraphics[width=0.2857142857142857\linewidth]{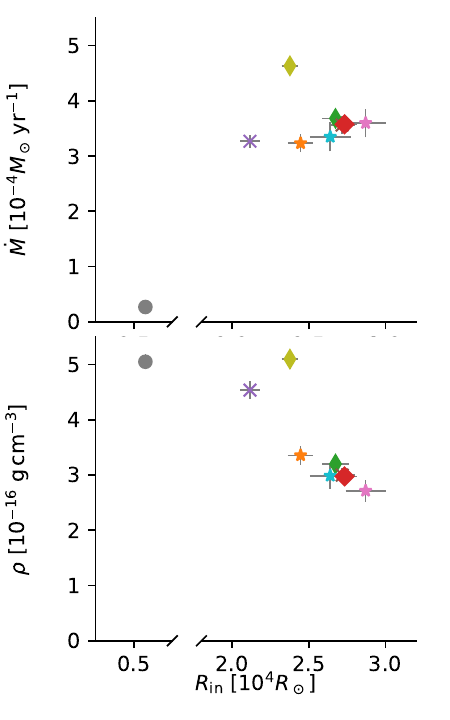}
\includegraphics[width=0.2285714285714285\linewidth]{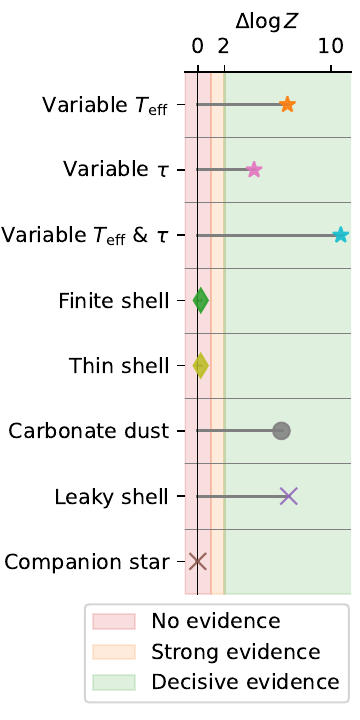}
\caption{
The progenitor properties of SN~2023ixf predicted by our baseline and alternative models.
The left panel shows $\Teff$ and $\logL$ of the best-fitting baseline model (red diamond) and alternative models (same symbols as in the right panel), compared to the SN II progenitors summarized in \citet{Smartt15}. MIST stellar evolution tracks \citep{Choi16} of different masses are overlaid for comparison.
The middle panel shows the inner radius of the dust envelope ($R_\text{in}$), the mass loss rate ($\dot{M}$, upper), and the average CSM density inside $R_\text{in}$ ($\rho$, lower) of the progenitor, derived from the baseline and alternative models, using the same symbols.
Finally, the right panel shows the increase of Bayesian evidence ($\log Z$) for alternative models compared to the baseline model. 
}
\label{fig:fig4}
\end{figure*}

\subsection{Progenitor identification and forced photometry}

To confirm the spatial coincidence of SN~2023ixf with the red source in the pre-explosion {\it HST}/ACS colour composite image, we first identify sources in the three-band combined {\it HST}/ACS image.
We use the iteratively-subtracted PSF photometry technique, which is optimized for crowded-field photometry, for this purpose.
First, inside a radius of $20$ arcsec to the SN, we choose 45 relatively isolated, high SNR stars and construct an effective PSF (EPSF) using \texttt{EPSFBulder} in \textsc{photutils}.
We then identify point-like sources with \texttt{DAOFind}, fit sources with the PSF, and subtract the best-fitting PSF from the image. The residual image is then used for another round of source detection and subtraction, and newly detected sources are fit together with sources detected in previous rounds.
We repeat this procedure until no new source is detectable in the residual image at a SNR threshold of $2$.
Based on the residual images, the sensitivity limit of the source detection procedure is better than $m_\text{F435W}\simeq 27.4$, $m_\text{F555W}\simeq 27.0$, and $m_\text{F814W}\simeq 27.4 $, assuming that the three-band combined image increases the sensitivity.

The red source is detected at a high SNR of $15.9$ with a position uncertainty of $3.3$~\mas. The distance of the SN to this source is $12.7\pm19.8$~\mas, consistent with spatial coincidence. There is another fainter source near the red source detected at a SNR of $6.7$ during the second round of iteration, with a position uncertainty of $10.3$~\mas. The distance of this source to the SN is $128.9\pm22.1$~\mas, which clearly rules out the possibility of spatial coincidence.
Based on the distances, we believe that the red source is the progenitor of SN~2023ixf.
We confirm the progenitor candidate identified in earlier works (e.g., \citealt{Pledger23}; \citetalias{Kilpatrick23}; \citealt{VanDyk23_23ixf}) with a substantially improved level of accuracy.
However, it should be emphasized that the identification must be double-checked with future revisiting after the SN has faded out when the progenitor is expected to have disappeared (e.g., \citealt{VanDyk23}).

Since the progenitor resides in a relatively crowded field with a non-smooth background contributed by unresolved sources, we use PSF photometry to measure the source flux in the calibrated science images, with pixel area map and charge transfer efficiency corrections applied, if available.
We choose the EPSF models of \citet{Anderson00, Anderson06} and \citet{Bellini18}, linearly interpolated at the detector position of the progenitor. For focus position-dependent EPSF models, we choose the focal distance that best minimizes the median flux uncertainty of the sources. In case no EPSF is provided for a specific filter, we choose the one with the nearest pivotal wavelength for the same instrument.
We make $12$ arcsec square cutouts centred at the progenitor and place EPSF models at the source positions detected earlier, including those that are not identified as the progenitor.
The source fluxes are fit as free parameters with their positions fixed, and the photometric calibration is based on the zero points in the FITS header.
We calculate the inverse variance-weighted average source flux for each unique combination of instrument, filter, and epoch of observation.
The averaged flux values are corrected for extinction assuming a Galactic reddening of $E(B-V)_\text{\,MW}=0.08$ mag \citep{SFD98, Schlafly11}, host galaxy reddening of $E(B-V)_\text{\,host}=0.031$ mag \citep{Smith23}, and the extinction coefficients of \citet{Schlafly11} with $R_\text{V}=3.1$.
The measured flux and magnitude (or 2$\sigma$ limiting magnitude, based on the forced-photometry flux uncertainty alone) are summarized in Table \ref{tab:hst_phot}.

\subsection{Progenitor physical properties}
\label{sec:properties}

To constrain the properties of the progenitor star and its CSM, we fit the SED of a variable dusty RSG to our {\it HST} optical and literature-compiled infrared photometry.

We generate a grid of SEDs with stellar effective temperature ($\Teff$) and dust optical depth ($\tau$) at $1\mu\text{m}$ as parameters.
The spectrum of the central star is based on the Model Atmospheres with a Radiative and Convective Scheme (MARCS; \citealt{Gustafsson08}) stellar atmosphere models.
We use the spectra of a solar-abundance massive giant ($15\,\mathrm{M}_\odot$, $\log [g/(\text{cm}\,\text{s}^{-2})]=0$) to cover the $\Teff$ range of $3400$ to $4000$ K; we further extend the $\Teff$ coverage down to $2400$ K using the spectra of a $5\,\mathrm{M}_\odot$ star.\footnote{We assume a solar-metallicity progenitor here, but there is tentative evidence for a sub-solar metallicity near the SN site ($\sim 0.7\,Z_\odot$; \citealt{Niu23, VanDyk23_23ixf, Zimmerman23}).}
The dust optical depth covers the range of $0.001$ to $50$, in logarithmic spacing.
The model grid includes $14$ nodes along the $\Teff$ axis and $50$ nodes along the $\tau$ axis.
We use \textsc{dusty} \citep{Ivezic97} for dust radiative transfer modelling, with a similar setup as described in \citet{Villaume15}, including both oxygen-rich and carbon-rich dust compositions.
The circumstellar dust has a $r^{-2}$ density profile, where the inner radius ($R_\text{in}$) is related to the dust condensation temperature (fixed at $1100$ K for carbon-rich and $700$ K for oxygen-rich compositions) and stellar luminosity, while the ratio of the outer-to-inner radius is fixed at $10^3$, representing an extended dust envelope.
We normalize the output SED to unit bolometric flux for interpolation and rescaling.

We construct a model with bolometric flux (in $\log F$), stellar effective temperature ($\Teff$), and circumstellar dust optical depth (in $\log \tau$) as free model parameters.
Given the strong, semi-regular variability of the progenitor \citepalias{Jencson23, Kilpatrick23, Soraisam23}, 
we allow stellar and dust physical parameters ($\log F$, $\Teff$, and $\log\tau$) to vary in a sinusoidal pattern with a regular period of $P$ (a free model parameter) spanning the time frame of our photometric dataset; the amplitude and initial phase of each parameter are also free model parameters.
To calculate the flux density in a specific band and epoch of observation, we interpolate the grid based on the $\Teff$ and $\log\tau$, scale the interpolated SED by the bolometric flux ($F$), and calculate the average flux density, weighted by the filter transmission profile obtained from the SVO Filter Profile Service.\footnote{\url{http://svo2.cab.inta-csic.es/theory/fps/}}
We fit the measured flux densities even in the absence of a statistically significant detection.
We use \textsc{emcee} for MCMC sampling, while the best-fitting parameters are the peaks of 1-d marginalized posterior distributions.
The goodness of fit is evaluated by the Bayesian evidence ($\log Z$) calculated using \textsc{dynesty} \citep{Speagle20, Koposov23}, a package for dynamic nested sampling \citep{Higson19}.

We choose the model with constant $\Teff$ and $\log\tau$, variable $\log F$, and oxygen-rich dust composition as the baseline model for comparison. The best-fitting model has a temperature of $\Teff=3342\pm27$ K and a phase-averaged bolometric luminosity of $\logL=5.10\pm0.02$ (including the uncertainty in the host galaxy distance), placing the star at the luminous side of the SN II progenitors population (Figure \ref{fig:fig4}, left).
The best-fitting luminosity and temperature imply a stellar radius of $R_\star=(1.06\pm0.03)\times10^3\,\mathrm{R}_\odot$ --- a greater radius compared to some of the largest known SN II progenitors (e.g., $R_\star\sim 740\,\mathrm{R}_\odot$ in \citealt{Soumagnac20}).
Based on the interpolated MIST stellar evolution tracks \citep{Choi16}, the best-fitting values and error ellipse of $\Teff$ and $\logL$ correspond to a solar-metallicity, post-main sequence star of $18.1_{-1.2}^{+0.7}\,\mathrm{M}_\odot$, also at the massive end of the SN II progenitor population.
The progenitor bolometric luminosity varies with an amplitude of $\Delta \log (L/\mathrm{L}_\odot)=0.11\pm0.01$ (i.e., $25.2\pm1.7$ per cent variation around the mean, or a peak-to-valley ratio of $1.66$) over a period of $P=1128.3\pm6.5$ days.
This indicates a factor of $1.29\pm 0.03$ change in the stellar radius from minimum to peak, assuming a constant $\Teff$.
At the time of the explosion ($85.8\pm1.1$ days after the last maximum), the progenitor luminosity is $\logL\sim5.20$, with a radius of $R_\star\sim 1.19\times 10^3\,\mathrm{R}_\odot$.

\begin{table}
\centering
\caption{Forced PSF Photometry of the Progenitor in Archival {\it HST} Images}
\label{tab:hst_phot}
\begin{tabular}{ccccc} 
\hline
Filter & MJD & Exposure [s] & Flux [$\mu$Jy] & Magnitude            \\ \hline

\multicolumn{5}{c}{Prop. ID 6829 (PI: You-Hua Chu), WFPC2}          \\ \hline
F656N & 51260.98 & 1360 & -0.410$\pm$3.777 & $>21.70$               \\
F675W & 51261.05 & 900 & 0.130$\pm$0.053 & $26.12_{-0.37}^{+0.57}$  \\
F547M & 51261.15 & 1400 & -0.011$\pm$0.057 & $>26.26$               \\
F656N & 51345.99 & 1200 & 0.228$\pm$10.37 & $>20.61$                \\
F547M & 51346.06 & 1000 & 0.123$\pm$0.331 & $>24.35$                \\[0.2cm]

\multicolumn{5}{c}{Prop. ID 9490 (PI: Kip Kuntz), ACS/WFC}          \\ \hline
F435W & 52594.00 & 900 & 0.008$\pm$0.010 & $>28.09$                 \\
F555W & 52594.01 & 720 & 0.006$\pm$0.013 & $>27.87$                 \\
F814W & 52594.02 & 720 & 0.378$\pm$0.016 & $24.96_{-0.04}^{+0.05}$  \\[0.2cm]

\multicolumn{5}{c}{Prop. ID 9720 (PI: Rupali Chandar), WFPC2}       \\ \hline
F336W & 52878.33 & 2400 & 0.096$\pm$0.890 & $>23.27$                \\[0.2cm]

\multicolumn{5}{c}{Prop. ID 13361 (PI: William Blair), WFC3/UVIS}   \\ \hline
F502N & 56735.86 & 1310 & 0.146$\pm$0.124 & $>25.42$                \\
F673N & 56735.87 & 1310 & 0.181$\pm$0.082 & $25.76_{-0.41}^{+0.66}$ \\[0.2cm]

\multicolumn{5}{c}{Prop. ID 15192 (PI: Benjamin Shappee), ACS/WFC}  \\ \hline
F658N & 58207.54 & 2956 & -0.100$\pm$0.053 & $>26.33$               \\
F435W & 58207.56 & 3712 & 0.007$\pm$0.007 & $>28.50$                \\ \hline
\end{tabular}
\end{table}

The best-fitting optical depth of the dust envelope is $\tau=2.83\pm0.03$ for the baseline model, which translates to an optical-band extinction of $A_\text{V}=10.28\pm0.11$ mag.
Such a circumstellar dust extinction is extremely heavy; only one RSG in the sample of \citet{Beasor22} has a comparably high $A_\text{V}$.
Assuming the opacity in \citet{Villaume15}, a gas-to-dust mass ratio of $\delta=200$ for solar-metallicity giants (e.g., \citealt{Mauron11, vanLoon05}), and a speed of $v_\text{wind}=50\,\kms$ for dust-driven wind, the optical depth indicates a mass-loss rate of $\dot{M}=(3.58\pm0.15) \times 10^{-4}\,\Msunyr$. 
Such a mass-loss rate is substantially higher than the empirical mass-loss rates of stars with similar $\logL$ and $\Teff$, for example, 
$[0.91_{-0.21}^{+0.27}\text{(stat.)}_{-0.57}^{+1.58}\text{(sys.)}]\times 10^{-5}\,\Msunyr$ assuming the \citet{Nieuwenhuijzen90} relationship for general stars,
or $[4.25_{-0.26}^{+0.28}\text{(stat.)}_{-2.12}^{+4.23}\text{(sys.)}]\times 10^{-5}\,\Msunyr$ assuming the \citet{vanLoon05} relationship for dusty RSGs and Asymptotic Giant Branch (AGB) stars.
However, it is in line with the period-dependent empirical mass-loss rate in \citet{Goldman17} for AGB stars and RSGs ($[2.02_{-0.68}^{+0.67}\text{(stat.)}_{-1.52}^{+6.66}\text{(sys.)}]\times 10^{-4}$ $\Msunyr$), given the large scatter of the relationship.

The luminosity and optical depth indicate a dust envelope inner radius of $R_\text{in}=(2.73\pm0.08)\times10^4\,\mathrm{R}_\odot$, within which the average CSM density is $\rho=(2.98\pm0.13)\times 10^{-16} \,\text{g}\,\text{cm}^{-3}$, close to the estimate in \citet{Zimmerman23} at the same radius in the extended wind region.
Inside the radius of $R_\text{p}=v_\text{wind}P=7.0\times 10^3\,\mathrm{R}_\odot$, i.e., the distance that stellar wind travels during one period of radial pulsation, or ``shell of pulsational mass loss,'' the average CSM density is $\rho=(4.55\pm0.20)\times 10^{-15}\,\text{g}\,\text{cm}^{-3}$, also close to the extended wind density outside the shock breakout radius in \citet{Zimmerman23}.
Notably, $R_\text{p}$ is close to the radial extension of the confined CSM ($R_\text{CSM}$) traced by the vanishing narrow emission lines.
Assuming a shock velocity of $v_\text{s}=10^4\,\kms$ and a timescale of $5$ days for the observed narrow emission lines, i.e., the time that the shock propagates within the dense CSM where efficient Compton cooling produces the ionizing radiation in extreme ultraviolet, the CSM radius is at most $R_\text{CSM}\simeq 6.2\times10^3\,\mathrm{R}_\odot$.
The coincidence of $R_\text{CSM}$ and $R_\text{p}$ implies that the dense, confined CSM is ejected during, but not necessarily driven by, the final episode of radial pulsation.
The CSM radius in \citet{Zimmerman23} is $2.9\times 10^3\,\mathrm{R}_\odot$, about half the simple estimate with $v_\text{s}$ above. As the progenitor exploded near its peak luminosity, the dense CSM may have been ejected around the minimum, about half a period before the explosion.
However, the complex structure of the shocked region and the increasing light crossing time introduce uncertainties in the shock propagation time in the dense CSM. The wind and shock speed also bear uncertainties (e.g., $v_\text{wind}=115\,\kms$ in \citealt{Smith23}). Therefore, $R_\text{CSM}$ and $R_\text{p}$ could differ by up to a factor of few.
Moreover, the estimated CSM density based on post-explosion observations indicates a substantially higher $\dot{M}$ compared to estimates from progenitor properties ($\sim 10^{-2}\,\Msunyr$ in \citealt{JacobsonGalan23, Hiramatsu23, Zimmerman23}), even higher than the typical mass-loss rates under the ``superwind'' scenario (e.g., \citealt{Forster18}), which requires a different mass-loss mechanism than dust-driven and pulsation-enhanced stellar winds.

We then compare the baseline model with a series of alternative models using the improvement in the Bayesian evidence ($\Delta \log Z$, i.e., the Bayes factor across two models), where a positive $\Delta \log Z$ indicates a more favourable model compared to the baseline model, given the existing dataset.
To calculate $\log Z$ efficiently, the period of variation is fixed at $P=1128.3$ days for the comparison here.
We choose a threshold of $\Delta \log Z = 2$ \citep{Jeffreys39} for decisive evidence in favour of an alternative model (Figure \ref{fig:fig4}, right).

First, we consider the scenario in which the variation of luminosity is accompanied by the variation of $\Teff$ or $\log\tau$ with the same period, characterized by their amplitudes and phase lags with respect to the variation of luminosity.
We find that either a periodic change in $\Teff$ (with an amplitude of $\Delta \Teff=563\pm64$ K and a phase lag of $354\pm40$ days, i.e., $\Teff$ peaks about $0.3$ periods after maximum light), or a change in the dust optical depth (with an amplitude of $\Delta \log \tau=0.059\pm0.015$ and phase lag of $506\pm36$ days, about half a period after maximum light) is more favourable compared to the baseline model.
Allowing both $\Teff$ and $\log\tau$ to vary, this more complex model better fits the data than the models in which only one varies, with similar amplitudes and phase lags in $\Teff$ and $\log\tau$ ($\Delta \Teff=625\pm91$ K with a phase lag of $308\pm32$ days, and $\Delta \log \tau=0.033\pm0.011$ with a phase lag of $615\pm61$ days).
These alternative models predict similar $\logL$, $\Teff$, $R_\text{in}$, and $\dot{M}$ as the baseline model (Figure \ref{fig:fig4}, left and middle).

We note that alternative models with a variable $\tau$ better reproduce the observed root mean square (RMS) amplitude-wavelength relationship in \citetalias{Soraisam23} (i.e., stronger variability towards shorter wavelengths) than the baseline model and the variable-$\Teff$ model (Figure \ref{fig:fig3}, upper right).
Therefore, the periodic variation of luminosity \emph{must} be accompanied by the variation of either $\Teff$ or $\log\tau$, if not both. Physically, this implies the change of stellar or dust properties and hence SED shape over the period.
Since $R_\text{in}$ is about $4$ times greater than $R_\text{p}$ and $25$ times greater than $R_\star$, instead of seeing the production of fresh dust during radial pulsations, the change in $\tau$ is likely due to the sublimation and condensation of dust out to a greater distance following the change of stellar irradiance.
The half-period phase lag in the variation of $\log\tau$ indicates that the dust column density (and hence mass) peaks when the progenitor shrinks to its minimum radius, while the one-third-period phase lag in the variation of $\Teff$ implies that the rate of dust condensation peaks after $\Teff$ begins to decrease.

Second, we consider the scenario in which the dust envelope has a finite radial extent, characterized by the ratio of outer to inner radius ($Y=R_\text{out}/R_\text{in}$).
Besides the baseline model which has a very extended, ``infinite'' dust envelope ($Y=10^3$), we consider the case of a finite ($Y=10$) and a thin ($Y=2$) dust shell.
We find that finite dust shells do not improve the goodness of fit ($\Delta\log Z=0.22$ for $Y=10$; $\Delta\log Z=0.01$ for $Y=2$), in contrast to the conclusion in \citet{Kilpatrick18} for the progenitor of SN~2017eaw.
The thin dust shell also leads to lower $\logL$ and $\Teff$ than the baseline model.
Even the optical depth (or dust column density) can be estimated, the outer radius of the dust envelope cannot be constrained by the dataset given the subtle differences in the resulting SED.

Third, the circumstellar dust of RSGs are mainly oxygen-rich silicates, but here we consider an alternative model with a carbon-rich composition.
We find that using carbonate dust improves the quality of fit ($\Delta Z=6.28$) than the oxygen-rich baseline model -- a similar conclusion as in \citet{Kilpatrick18} for SN~2017eaw and \citet{Niu23} for SN~2023ixf.
However, the carbonate dust model leads to a significantly cooler and lower luminosity progenitor, beyond the coverage of MIST isochrones (Figure \ref{fig:fig4}, left).
The implied dust envelope inner radius and mass-loss rate are also lower than models based on oxygen-rich silicate dust (Figure \ref{fig:fig4}, middle).
The assumed dust condensation temperature could have contributed to the systematic differences here.
We conclude that despite the improvement in the Bayesian evidence, using carbonate dust may lead to biased and even unphysical progenitor properties.
Furthermore, distinguishing dust composition requires photometry in the mid-infrared range, which is not covered by the dataset.

Fourth, the CSM around the progenitor is likely asymmetric \citep{Smith23, Vasylyev23}. Therefore, the progenitor could be partially or non-uniformly obscured by the circumstellar dust.
We consider the case in which a fraction of the progenitor's light has escaped without being absorbed and reemitted by the dust.
The best-fitting escape fraction of this ``leaky shell'' model is $f_\text{esc}=18.5\pm2.0$ per cent, with an increase in the Bayesian evidence of $\Delta\log Z=6.83$, indicating that either non-spherical or clumpy dust could better fit the observed SED. However, the model prefers a cooler star compared to the baseline model.
Despite the improvement in $\Delta\log Z$, we note that the wavelength coverage of our dataset might not be able to constrain the dust geometry effectively.

Finally, we consider the potential contribution of an unobscured binary companion star in the observed SED.
Assuming that the companion star lies on the same best-matching MIST isochrone as the progenitor, we use the companion ZAMS mass ($M_2$) as the free parameter with a flat prior.
We add a new SED component based on the BaSeL v3.1 stellar template \citep{Lejeune97} using the effective temperature and luminosity predicted from the isochrone.
The best-fitting model has $M_2=4.5\pm1.9\, \mathrm{M}_\odot$, close to a main sequence star with $T_\text{eff,2}=(1.59\pm0.41)\times10^{4}$ K and $\log (L_2/\mathrm{L}_\odot)=2.57\pm0.73$. The single-sided $95\%$ upper limit is $7.1\, \mathrm{M}_\odot$, or $\log (L_2/\mathrm{L}_\odot)=3.29$ in luminosity.
The companion star model does not outperform the baseline model ($\Delta\log Z=0.06$). Nevertheless, the sensitivity limit of our data may not confirm the single-star nature of the progenitor; only companions of $M_2>7.1\,\mathrm{M}_\odot$ can be robustly ruled out.

In Table \ref{tab:comp}, we summarize the progenitor properties in earlier works and our results.
We derive consistent $\logL$ and $M$ values compared to other works, except for \citetalias{Kilpatrick23}, which prefers a lower luminosity and hence a lower-mass progenitor.
\citetalias{Soraisam23} estimated a marginally higher luminosity based on the period-luminosity relationship in \citet{Soraisam18}; nevertheless, the estimated mass is consistent with our result.
The effective temperature is not robustly constrained in general; we find a $\Teff$ that is consistent with earlier works but cooler than \citetalias{Kilpatrick23}.
Furthermore, we find a comparable dust optical depth (or extinction) with \citet{VanDyk23_23ixf} but higher than other works. 
A higher $\tau$ value, along with the larger $R_\text{in}$ (e.g., compared to $8600\,\mathrm{R}_\odot$ in \citetalias{Kilpatrick23}), a dust temperature-sensitive property, leads to a higher $\dot{M}$.
Notably, the analyses in \citetalias{Jencson23} and \citet{VanDyk23_23ixf} are based on the Grid of RSG and AGB ModelS (GRAMS; \citealt{Sargent11, Srinivasan11}), while the key results are not systematically different than works using MARCS and \textsc{dusty} for SED modelling.
The variance across these independent analyses is attributable to the different modelling strategies (e.g., the choice of dust composition) and subsets of archival data used.

\renewcommand{\arraystretch}{1.25}
\begin{table*}
\centering
\caption{Key Progenitor Properties Compared to Other Works} \label{tab:comp}
\begin{tabular}{lccccccc}
\hline
                        & $\logL$                   & $\Teff$ [K]           & $M$ [$\mathrm{M}_\odot$]       & $\tau$ (1$\mu$m)                      & $A_\text{V}$ [mag]     & $\dot{M}$ [$10^{-4}\Msunyr$]                 & SED Model             \\ \hline
\citet{Jencson23}       & $5.1\pm0.2$               & $3500^{+800}_{-1400}$ & $17\pm4$              & $2.2$                                 & --                     & $1.5$ to $15^\text{\,a}$                     & Mainly GRAMS          \\
\citet{Kilpatrick23}    & $4.74\pm0.07$             & $3920^{+200}_{-160}$  & $\sim11$              & --                                    & $4.6\pm0.2$            & $0.026\pm0.002^\text{\,a}$                   & MARCS+\textsc{dusty}  \\
\citet{Niu23}           & $5.11\pm0.08$             & $3700^\text{\,b}$     & $16.2$ to $17.4$      & --                                    & $6.94^{+0.63}_{-0.64}$ & $\sim 0.43^\text{\,a}$                       & MARCS+\textsc{dusty}  \\
\citet{Pledger23}       & --                        & --                    & $8$ to $10$           & --                                    & --                     & --                                           & --                    \\
\citet{Soraisam23}      & $5.27\pm0.12$             & $3200^\text{\,b}$     & $20\pm4$              & --                                    & --                     & $2$ to $4^\text{\,d}$                        & --                    \\
$\cdots$                & $5.37\pm0.12$             & $3500^\text{\,b}$     & $\cdots$              & --                                    & --                     & $\cdots$                                     & --                    \\
\citet{VanDyk23_23ixf}  & $4.97_{-0.09}^{+0.06}$    & $3450_{-1080}^{+250}$ & $12$ to $15$          & $1.95_{-0.18}^{+0.96}$$^\text{\,e}$   & --                     & $0.104_{-0.022}^{+0.198}$ $^\text{\,e,f}$    & GRAMS                 \\ \hline
\textbf{This work}      & $5.10\pm0.02$             & $3343\pm27$           & $18.1_{-1.2}^{+0.7}$  & $2.83\pm0.03$         & $10.28\pm0.11$         & $3.58\pm0.15$                                & MARCS+\textsc{dusty}  \\ \hline
\multicolumn{8}{l}{$^\text{a}$ Scaled to $v_\text{wind}=50\kms$ and $\delta=200$. $^\text{b}$ Fixed parameter. $^\text{c}$ Based on the best-matching isochrone in the color-magnitude diagram.} \\
\multicolumn{8}{l}{$^\text{d}$ Inferred from period and luminosity \citep{Goldman17} assuming $\delta=200$.} \\
\multicolumn{8}{l}{$^\text{e}$ Based on the maximum-likelihood estimate and 16th, 84th percentiles. $^\text{f}$ Converted from the dust production rate assuming $\delta=200$.}
\end{tabular}
\end{table*}

\section{Summary and Discussion}

We identify the progenitor star of SN~2023ixf in the pre-explosion {\it HST}/ACS image using Keck/NIRC2 adaptive optics imaging.
The SN position, precisely determined to a total uncertainty of $19.5$~\mas, unambiguously coincides with a red source in the {\it HST}/ACS image; other sources, including a nearby source detected using iteratively-subtracted PSF photometry, are ruled out. With forced PSF photometry, we obtain $2\sigma$ detections of the progenitor in three {\it HST} bands.

Given the reported infrared excess and variability of the progenitor, we fit the SED of a dusty variable RSG to a combined dataset including our {\it HST} photometry and infrared measurements in the literature.
We find $\logL=5.10\pm0.02$ and $\Teff=3343\pm27$ K for the best-fitting model, consistent with a post-main sequence massive single star of $18.1_{-1.2}^{+0.7}\,\mathrm{M}_\odot$, among the most luminous and massive SN II progenitors.
The heavy dust obscuration ($\tau=2.83\pm0.03$ at $1\,\mu\text{m}$) indicates an enhanced pre-SN mass loss rate of $(3.58\pm0.15)\times 10^{-4}\,\Msunyr$ and a CSM density of $(4.55\pm0.20)\times 10^{-15}\,\text{g}\,\text{cm}^{-3}$ inside the shell of pulsational mass loss.
Based on the timescale of the observed narrow emission lines and the period of progenitor variability, we suggest that the dense and confined CSM is ejected during the last episode of radial pulsation before the explosion.

We find strong evidence for the synchronized variation of dust or stellar properties along with the variation of luminosity.
Specifically, alternative models with a variable dust optical depth better reproduce the observed amplitude-wavelength relationship.
We suggest that the luminosity variation and radial pulsation of the progenitor may lead to periodic dust sublimation and condensation, and hence the change in $\tau$, near the inner radius of the dust envelope.
However, the change in other dust properties (e.g., temperature, grain size) could also lead to the apparent variability of $\tau$.

Furthermore, non-spherical dust geometry or partial dust obscuration remains possible; about $18.5\pm1.2$ per cent of the progenitor's light may have escaped without being reprocessed by the circumstellar dust envelope. However, any companion star above $7.1\,\mathrm{M}_\odot$ can be ruled out based on the dataset.

We conclude that the progenitor of SN~2023ixf is among the most massive, luminous, and heavily obscured SN II progenitors, which likely experienced enhanced mass loss before the explosion.

\section*{Acknowledgements}


The authors would like to thank Jacob Jencson for kindly providing his {\it Spitzer}/IRAC and MMIRS measurements before the acceptance of \citet{Jencson23}. The authors would also like to thank Ping Chen, Ningchen Sun, and Subo Dong for their valuable comments on this work. 
YQ thanks Jianwei Lyu and Fengwu Sun for the discussion on infrared flux calibration, and Lile Wang for the discussion on circumstellar dust thermodynamics.
YQ thanks Weidong Li (deceased), whose work on the progenitor of SN~2011fe in Messier 101 \citep{Li11_11fe} motivated his career as an astrophysicist.

AGY's research is supported by the EU via ERC grant No. 725161, the ISF GW excellence centre, an IMOS space infrastructure grant and BSF/Transformative and GIF grants, as well as the Andr\'{e} Deloro Institute for Advanced Research in Space and Optics, The Helen Kimmel Center for Planetary Science, the Schwartz/Reisman Collaborative Science Program and the Norman E Alexander Family M Foundation ULTRASAT Data Center Fund, Minerva and Yeda-Sela; AGY is the incumbent of the The Arlyn Imberman Professorial Chair.
MWC acknowledges support from the National Science Foundation with grant numbers PHY-2010970 and OAC-2117997.

SS acknowledges support from the G.R.E.A.T. research environment, funded by {\em Vetenskapsr\aa det},  the Swedish Research Council, project number 2016-06012.

\section*{Data Availability}

The compiled photometric dataset, SED model grid, and the best-fitting result are available upon request.
 



\bibliographystyle{mnras}
\bibliography{main} 

\begin{thebibliography}{}
\makeatletter
\relax
\def\mn@urlcharsother{\let\do\@makeother \do\$\do\&\do\#\do\^\do\_\do\%\do\~}
\def\mn@doi{\begingroup\mn@urlcharsother \@ifnextchar [ {\mn@doi@} {\mn@doi@[]}}
\def\mn@doi@[#1]#2{\def\@tempa{#1}\ifx\@tempa\@empty \href {http://dx.doi.org/#2} {doi:#2}\else \href {http://dx.doi.org/#2} {#1}\fi \endgroup}
\def\mn@eprint#1#2{\mn@eprint@#1:#2::\@nil}
\def\mn@eprint@arXiv#1{\href {http://arxiv.org/abs/#1} {{\tt arXiv:#1}}}
\def\mn@eprint@dblp#1{\href {http://dblp.uni-trier.de/rec/bibtex/#1.xml} {dblp:#1}}
\def\mn@eprint@#1:#2:#3:#4\@nil{\def\@tempa {#1}\def\@tempb {#2}\def\@tempc {#3}\ifx \@tempc \@empty \let \@tempc \@tempb \let \@tempb \@tempa \fi \ifx \@tempb \@empty \def\@tempb {arXiv}\fi \@ifundefined {mn@eprint@\@tempb}{\@tempb:\@tempc}{\expandafter \expandafter \csname mn@eprint@\@tempb\endcsname \expandafter{\@tempc}}}

\bibitem[\protect\citeauthoryear{{Anderson} \& {King}}{{Anderson} \& {King}}{2000}]{Anderson00}
{Anderson} J.,  {King} I.~R.,  2000, \mn@doi [\pasp] {10.1086/316632}, \href {https://ui.adsabs.harvard.edu/abs/2000PASP..112.1360A} {112, 1360}

\bibitem[\protect\citeauthoryear{{Anderson} \& {King}}{{Anderson} \& {King}}{2006}]{Anderson06}
{Anderson} J.,  {King} I.~R.,  2006, {PSFs, Photometry, and Astronomy for the ACS/WFC}, Instrument Science Report ACS 2006-01, 34 pages

\bibitem[\protect\citeauthoryear{{Beasor} \& {Davies}}{{Beasor} \& {Davies}}{2016}]{Beasor16}
{Beasor} E.~R.,  {Davies} B.,  2016, \mn@doi [\mnras] {10.1093/mnras/stw2054}, \href {https://ui.adsabs.harvard.edu/abs/2016MNRAS.463.1269B} {463, 1269}

\bibitem[\protect\citeauthoryear{{Beasor} \& {Smith}}{{Beasor} \& {Smith}}{2022}]{Beasor22}
{Beasor} E.~R.,  {Smith} N.,  2022, \mn@doi [\apj] {10.3847/1538-4357/ac6dcf}, \href {https://ui.adsabs.harvard.edu/abs/2022ApJ...933...41B} {933, 41}

\bibitem[\protect\citeauthoryear{{Bellini}, {Anderson}  \& {Grogin}}{{Bellini} et~al.}{2018}]{Bellini18}
{Bellini} A.,  {Anderson} J.,   {Grogin} N.~A.,  2018, {Focus-diverse, empirical PSF models for the ACS/WFC}, Instrument Science Report ACS 2018-8

\bibitem[\protect\citeauthoryear{{Berger} et~al.,}{{Berger} et~al.}{2023}]{Berger23}
{Berger} E.,  et~al., 2023, \mn@doi [\apjl] {10.3847/2041-8213/ace0c4}, \href {https://ui.adsabs.harvard.edu/abs/2023ApJ...951L..31B} {951, L31}

\bibitem[\protect\citeauthoryear{{Bianciardi} et~al.,}{{Bianciardi} et~al.}{2023}]{Bianciardi23}
{Bianciardi} G.,  et~al., 2023, \mn@doi [Transient Name Server AstroNote] {10.48550/arXiv.2307.05612}, \href {https://ui.adsabs.harvard.edu/abs/2023TNSAN.213....1B} {213, 1}

\bibitem[\protect\citeauthoryear{{Bostroem} et~al.,}{{Bostroem} et~al.}{2023}]{Bostroem23}
{Bostroem} K.~A.,  et~al., 2023, \mn@doi [arXiv e-prints] {10.48550/arXiv.2306.10119}, \href {https://ui.adsabs.harvard.edu/abs/2023arXiv230610119B} {p. arXiv:2306.10119}

\bibitem[\protect\citeauthoryear{Bradley et~al.,}{Bradley et~al.}{2022}]{Bradley22}
Bradley L.,  et~al., 2022, astropy/photutils: 1.5.0, \mn@doi{10.5281/zenodo.6825092}, \url {https://doi.org/10.5281/zenodo.6825092}

\bibitem[\protect\citeauthoryear{{Choi}, {Dotter}, {Conroy}, {Cantiello}, {Paxton}  \& {Johnson}}{{Choi} et~al.}{2016}]{Choi16}
{Choi} J.,  {Dotter} A.,  {Conroy} C.,  {Cantiello} M.,  {Paxton} B.,   {Johnson} B.~D.,  2016, \mn@doi [\apj] {10.3847/0004-637X/823/2/102}, \href {https://ui.adsabs.harvard.edu/abs/2016ApJ...823..102C} {823, 102}

\bibitem[\protect\citeauthoryear{{Davies} \& {Beasor}}{{Davies} \& {Beasor}}{2018}]{Davies18}
{Davies} B.,  {Beasor} E.~R.,  2018, \mn@doi [\mnras] {10.1093/mnras/stx2734}, \href {https://ui.adsabs.harvard.edu/abs/2018MNRAS.474.2116D} {474, 2116}

\bibitem[\protect\citeauthoryear{{Davies} \& {Beasor}}{{Davies} \& {Beasor}}{2020}]{Davies20}
{Davies} B.,  {Beasor} E.~R.,  2020, \mn@doi [\mnras] {10.1093/mnras/staa174}, \href {https://ui.adsabs.harvard.edu/abs/2020MNRAS.493..468D} {493, 468}

\bibitem[\protect\citeauthoryear{{Dong} et~al.,}{{Dong} et~al.}{2023}]{Dong23}
{Dong} Y.,  et~al., 2023, \mn@doi [arXiv e-prints] {10.48550/arXiv.2307.02539}, \href {https://ui.adsabs.harvard.edu/abs/2023arXiv230702539D} {p. arXiv:2307.02539}

\bibitem[\protect\citeauthoryear{{Flinner}, {Tucker}, {Beacom}  \& {Shappee}}{{Flinner} et~al.}{2023}]{Flinner23}
{Flinner} N.,  {Tucker} M.~A.,  {Beacom} J.~F.,   {Shappee} B.~J.,  2023, \mn@doi [Research Notes of the American Astronomical Society] {10.3847/2515-5172/acefc4}, \href {https://ui.adsabs.harvard.edu/abs/2023RNAAS...7..174F} {7, 174}

\bibitem[\protect\citeauthoryear{{Foreman-Mackey}, {Hogg}, {Lang}  \& {Goodman}}{{Foreman-Mackey} et~al.}{2013}]{ForemanMackey13}
{Foreman-Mackey} D.,  {Hogg} D.~W.,  {Lang} D.,   {Goodman} J.,  2013, \mn@doi [\pasp] {10.1086/670067}, \href {https://ui.adsabs.harvard.edu/abs/2013PASP..125..306F} {125, 306}

\bibitem[\protect\citeauthoryear{{F{\"o}rster} et~al.,}{{F{\"o}rster} et~al.}{2018}]{Forster18}
{F{\"o}rster} F.,  et~al., 2018, \mn@doi [Nature Astronomy] {10.1038/s41550-018-0563-4}, \href {https://ui.adsabs.harvard.edu/abs/2018NatAs...2..808F} {2, 808}

\bibitem[\protect\citeauthoryear{{Gaia Collaboration} et~al.,}{{Gaia Collaboration} et~al.}{2018}]{Gaia18}
{Gaia Collaboration} et~al., 2018, \mn@doi [\aap] {10.1051/0004-6361/201833051}, \href {https://ui.adsabs.harvard.edu/abs/2018A&A...616A...1G} {616, A1}

\bibitem[\protect\citeauthoryear{{Goldman} et~al.,}{{Goldman} et~al.}{2017}]{Goldman17}
{Goldman} S.~R.,  et~al., 2017, \mn@doi [\mnras] {10.1093/mnras/stw2708}, \href {https://ui.adsabs.harvard.edu/abs/2017MNRAS.465..403G} {465, 403}

\bibitem[\protect\citeauthoryear{{Grefenstette}, {Brightman}, {Earnshaw}, {Harrison}  \& {Margutti}}{{Grefenstette} et~al.}{2023}]{Grefenstette23}
{Grefenstette} B.~W.,  {Brightman} M.,  {Earnshaw} H.~P.,  {Harrison} F.~A.,   {Margutti} R.,  2023, \mn@doi [arXiv e-prints] {10.48550/arXiv.2306.04827}, \href {https://ui.adsabs.harvard.edu/abs/2023arXiv230604827G} {p. arXiv:2306.04827}

\bibitem[\protect\citeauthoryear{{Guetta}, {Langella}, {Gagliardini}  \& {Della Valle}}{{Guetta} et~al.}{2023}]{Guetta23}
{Guetta} D.,  {Langella} A.,  {Gagliardini} S.,   {Della Valle} M.,  2023, \mn@doi [arXiv e-prints] {10.48550/arXiv.2306.14717}, \href {https://ui.adsabs.harvard.edu/abs/2023arXiv230614717G} {p. arXiv:2306.14717}

\bibitem[\protect\citeauthoryear{{Gustafsson}, {Edvardsson}, {Eriksson}, {J{\o}rgensen}, {Nordlund}  \& {Plez}}{{Gustafsson} et~al.}{2008}]{Gustafsson08}
{Gustafsson} B.,  {Edvardsson} B.,  {Eriksson} K.,  {J{\o}rgensen} U.~G.,  {Nordlund} {\r{A}}.,   {Plez} B.,  2008, \mn@doi [\aap] {10.1051/0004-6361:200809724}, \href {https://ui.adsabs.harvard.edu/abs/2008A&A...486..951G} {486, 951}

\bibitem[\protect\citeauthoryear{{Higson}, {Handley}, {Hobson}  \& {Lasenby}}{{Higson} et~al.}{2019}]{Higson19}
{Higson} E.,  {Handley} W.,  {Hobson} M.,   {Lasenby} A.,  2019, \mn@doi [Statistics and Computing] {10.1007/s11222-018-9844-0}, \href {https://ui.adsabs.harvard.edu/abs/2019S&C....29..891H} {29, 891}

\bibitem[\protect\citeauthoryear{{Hiramatsu} et~al.,}{{Hiramatsu} et~al.}{2023}]{Hiramatsu23}
{Hiramatsu} D.,  et~al., 2023, \mn@doi [arXiv e-prints] {10.48550/arXiv.2307.03165}, \href {https://ui.adsabs.harvard.edu/abs/2023arXiv230703165H} {p. arXiv:2307.03165}

\bibitem[\protect\citeauthoryear{{Horiuchi}, {Nakamura}, {Takiwaki}, {Kotake}  \& {Tanaka}}{{Horiuchi} et~al.}{2014}]{Horiuchi14}
{Horiuchi} S.,  {Nakamura} K.,  {Takiwaki} T.,  {Kotake} K.,   {Tanaka} M.,  2014, \mn@doi [\mnras] {10.1093/mnrasl/slu146}, \href {https://ui.adsabs.harvard.edu/abs/2014MNRAS.445L..99H} {445, L99}

\bibitem[\protect\citeauthoryear{{Hosseinzadeh} et~al.,}{{Hosseinzadeh} et~al.}{2023}]{Hosseinzadeh23}
{Hosseinzadeh} G.,  et~al., 2023, \mn@doi [arXiv e-prints] {10.48550/arXiv.2306.06097}, \href {https://ui.adsabs.harvard.edu/abs/2023arXiv230606097H} {p. arXiv:2306.06097}

\bibitem[\protect\citeauthoryear{{Itagaki}}{{Itagaki}}{2023}]{Itagaki23TNS}
{Itagaki} K.,  2023, Transient Name Server Discovery Report, \href {https://ui.adsabs.harvard.edu/abs/2023TNSTR1158....1I} {2023-1158, 1}

\bibitem[\protect\citeauthoryear{{Ivezic} \& {Elitzur}}{{Ivezic} \& {Elitzur}}{1997}]{Ivezic97}
{Ivezic} Z.,  {Elitzur} M.,  1997, \mn@doi [\mnras] {10.1093/mnras/287.4.799}, \href {https://ui.adsabs.harvard.edu/abs/1997MNRAS.287..799I} {287, 799}

\bibitem[\protect\citeauthoryear{{Jacobson-Galan} et~al.,}{{Jacobson-Galan} et~al.}{2023}]{JacobsonGalan23}
{Jacobson-Galan} W.~V.,  et~al., 2023, \mn@doi [arXiv e-prints] {10.48550/arXiv.2306.04721}, \href {https://ui.adsabs.harvard.edu/abs/2023arXiv230604721J} {p. arXiv:2306.04721}

\bibitem[\protect\citeauthoryear{{Jeffreys}}{{Jeffreys}}{1939}]{Jeffreys39}
{Jeffreys} H.,  1939, {Theory of Probability}

\bibitem[\protect\citeauthoryear{{Jencson} et~al.,}{{Jencson} et~al.}{2023}]{Jencson23}
{Jencson} J.~E.,  et~al., 2023, \mn@doi [arXiv e-prints] {10.48550/arXiv.2306.08678}, \href {https://ui.adsabs.harvard.edu/abs/2023arXiv230608678J} {p. arXiv:2306.08678}

\bibitem[\protect\citeauthoryear{{Kilpatrick} \& {Foley}}{{Kilpatrick} \& {Foley}}{2018}]{Kilpatrick18}
{Kilpatrick} C.~D.,  {Foley} R.~J.,  2018, \mn@doi [\mnras] {10.1093/mnras/sty2435}, \href {https://ui.adsabs.harvard.edu/abs/2018MNRAS.481.2536K} {481, 2536}

\bibitem[\protect\citeauthoryear{{Kilpatrick} et~al.,}{{Kilpatrick} et~al.}{2023}]{Kilpatrick23}
{Kilpatrick} C.~D.,  et~al., 2023, \mn@doi [arXiv e-prints] {10.48550/arXiv.2306.04722}, \href {https://ui.adsabs.harvard.edu/abs/2023arXiv230604722K} {p. arXiv:2306.04722}

\bibitem[\protect\citeauthoryear{Koposov et~al.,}{Koposov et~al.}{2023}]{Koposov23}
Koposov S.,  et~al., 2023, joshspeagle/dynesty: v2.1.2, \mn@doi{10.5281/zenodo.7995596}, \url {https://doi.org/10.5281/zenodo.7995596}

\bibitem[\protect\citeauthoryear{{Lejeune}, {Cuisinier}  \& {Buser}}{{Lejeune} et~al.}{1997}]{Lejeune97}
{Lejeune} T.,  {Cuisinier} F.,   {Buser} R.,  1997, \mn@doi [\aaps] {10.1051/aas:1997373}, \href {https://ui.adsabs.harvard.edu/abs/1997A&AS..125..229L} {125, 229}

\bibitem[\protect\citeauthoryear{{Li} et~al.,}{{Li} et~al.}{2011a}]{Li11}
{Li} W.,  et~al., 2011a, \mn@doi [\mnras] {10.1111/j.1365-2966.2011.18160.x}, \href {https://ui.adsabs.harvard.edu/abs/2011MNRAS.412.1441L} {412, 1441}

\bibitem[\protect\citeauthoryear{{Li} et~al.,}{{Li} et~al.}{2011b}]{Li11_11fe}
{Li} W.,  et~al., 2011b, \mn@doi [\nat] {10.1038/nature10646}, \href {https://ui.adsabs.harvard.edu/abs/2011Natur.480..348L} {480, 348}

\bibitem[\protect\citeauthoryear{{Mao} et~al.,}{{Mao} et~al.}{2023}]{Mao23TNS}
{Mao} Y.,  et~al., 2023, Transient Name Server AstroNote, \href {https://ui.adsabs.harvard.edu/abs/2023TNSAN.130....1M} {130, 1}

\bibitem[\protect\citeauthoryear{{Mauron} \& {Josselin}}{{Mauron} \& {Josselin}}{2011}]{Mauron11}
{Mauron} N.,  {Josselin} E.,  2011, \mn@doi [\aap] {10.1051/0004-6361/201013993}, \href {https://ui.adsabs.harvard.edu/abs/2011A&A...526A.156M} {526, A156}

\bibitem[\protect\citeauthoryear{{M{\"u}ller}, {Carenza}, {Eckner}  \& {Goobar}}{{M{\"u}ller} et~al.}{2023}]{Muller23}
{M{\"u}ller} E.,  {Carenza} P.,  {Eckner} C.,   {Goobar} A.,  2023, \mn@doi [arXiv e-prints] {10.48550/arXiv.2306.16397}, \href {https://ui.adsabs.harvard.edu/abs/2023arXiv230616397M} {p. arXiv:2306.16397}

\bibitem[\protect\citeauthoryear{{Neustadt}, {Kochanek}  \& {Rizzo Smith}}{{Neustadt} et~al.}{2023}]{Neustadt23}
{Neustadt} J.~M.~M.,  {Kochanek} C.~S.,   {Rizzo Smith} M.,  2023, \mn@doi [arXiv e-prints] {10.48550/arXiv.2306.06162}, \href {https://ui.adsabs.harvard.edu/abs/2023arXiv230606162N} {p. arXiv:2306.06162}

\bibitem[\protect\citeauthoryear{{Nieuwenhuijzen} \& {de Jager}}{{Nieuwenhuijzen} \& {de Jager}}{1990}]{Nieuwenhuijzen90}
{Nieuwenhuijzen} H.,  {de Jager} C.,  1990, \aap, \href {https://ui.adsabs.harvard.edu/abs/1990A&A...231..134N} {231, 134}

\bibitem[\protect\citeauthoryear{{Niu}, {Sun}, {Maund}, {Zhang}, {Zhao}  \& {Liu}}{{Niu} et~al.}{2023}]{Niu23}
{Niu} Z.-X.,  {Sun} N.-C.,  {Maund} J.~R.,  {Zhang} Y.,  {Zhao} R.-N.,   {Liu} J.-F.,  2023, \mn@doi [arXiv e-prints] {10.48550/arXiv.2308.04677}, \href {https://ui.adsabs.harvard.edu/abs/2023arXiv230804677N} {p. arXiv:2308.04677}

\bibitem[\protect\citeauthoryear{{O'Connor} \& {Ott}}{{O'Connor} \& {Ott}}{2011}]{OConnor11}
{O'Connor} E.,  {Ott} C.~D.,  2011, \mn@doi [\apj] {10.1088/0004-637X/730/2/70}, \href {https://ui.adsabs.harvard.edu/abs/2011ApJ...730...70O} {730, 70}

\bibitem[\protect\citeauthoryear{{Panjkov}, {Auchettl}, {Shappee}, {Do}, {Lopez}  \& {Beacom}}{{Panjkov} et~al.}{2023}]{Panjkov23}
{Panjkov} S.,  {Auchettl} K.,  {Shappee} B.~J.,  {Do} A.,  {Lopez} L.~A.,   {Beacom} J.~F.,  2023, \mn@doi [arXiv e-prints] {10.48550/arXiv.2308.13101}, \href {https://ui.adsabs.harvard.edu/abs/2023arXiv230813101P} {p. arXiv:2308.13101}

\bibitem[\protect\citeauthoryear{{Perley}, {Gal-Yam}, {Irani}  \& {Zimmerman}}{{Perley} et~al.}{2023}]{Perley23TNS}
{Perley} D.~A.,  {Gal-Yam} A.,  {Irani} I.,   {Zimmerman} E.,  2023, Transient Name Server AstroNote, \href {https://ui.adsabs.harvard.edu/abs/2023TNSAN.119....1P} {119, 1}

\bibitem[\protect\citeauthoryear{{Pledger} \& {Shara}}{{Pledger} \& {Shara}}{2023}]{Pledger23}
{Pledger} J.~L.,  {Shara} M.~M.,  2023, \mn@doi [\apjl] {10.3847/2041-8213/ace88b}, \href {https://ui.adsabs.harvard.edu/abs/2023ApJ...953L..14P} {953, L14}

\bibitem[\protect\citeauthoryear{{Riess} et~al.,}{{Riess} et~al.}{2022}]{Riess22}
{Riess} A.~G.,  et~al., 2022, \mn@doi [\apjl] {10.3847/2041-8213/ac5c5b}, \href {https://ui.adsabs.harvard.edu/abs/2022ApJ...934L...7R} {934, L7}

\bibitem[\protect\citeauthoryear{{Sargent}, {Srinivasan}  \& {Meixner}}{{Sargent} et~al.}{2011}]{Sargent11}
{Sargent} B.~A.,  {Srinivasan} S.,   {Meixner} M.,  2011, \mn@doi [\apj] {10.1088/0004-637X/728/2/93}, \href {https://ui.adsabs.harvard.edu/abs/2011ApJ...728...93S} {728, 93}

\bibitem[\protect\citeauthoryear{{Sarmah}}{{Sarmah}}{2023}]{Sarmah23}
{Sarmah} P.,  2023, \mn@doi [arXiv e-prints] {10.48550/arXiv.2307.08744}, \href {https://ui.adsabs.harvard.edu/abs/2023arXiv230708744S} {p. arXiv:2307.08744}

\bibitem[\protect\citeauthoryear{{Schlafly} \& {Finkbeiner}}{{Schlafly} \& {Finkbeiner}}{2011}]{Schlafly11}
{Schlafly} E.~F.,  {Finkbeiner} D.~P.,  2011, \mn@doi [\apj] {10.1088/0004-637X/737/2/103}, \href {https://ui.adsabs.harvard.edu/abs/2011ApJ...737..103S} {737, 103}

\bibitem[\protect\citeauthoryear{{Schlegel}, {Finkbeiner}  \& {Davis}}{{Schlegel} et~al.}{1998}]{SFD98}
{Schlegel} D.~J.,  {Finkbeiner} D.~P.,   {Davis} M.,  1998, \mn@doi [\apj] {10.1086/305772}, \href {https://ui.adsabs.harvard.edu/abs/1998ApJ...500..525S} {500, 525}

\bibitem[\protect\citeauthoryear{{Sgro} et~al.,}{{Sgro} et~al.}{2023}]{Sgro23}
{Sgro} L.~A.,  et~al., 2023, \mn@doi [Research Notes of the American Astronomical Society] {10.3847/2515-5172/ace41f}, \href {https://ui.adsabs.harvard.edu/abs/2023RNAAS...7..141S} {7, 141}

\bibitem[\protect\citeauthoryear{{Shivvers} et~al.,}{{Shivvers} et~al.}{2017}]{Shivvers17}
{Shivvers} I.,  et~al., 2017, \mn@doi [\pasp] {10.1088/1538-3873/aa54a6}, \href {https://ui.adsabs.harvard.edu/abs/2017PASP..129e4201S} {129, 054201}

\bibitem[\protect\citeauthoryear{{Skrutskie} et~al.,}{{Skrutskie} et~al.}{2006}]{Skrutskie06}
{Skrutskie} M.~F.,  et~al., 2006, \mn@doi [\aj] {10.1086/498708}, \href {https://ui.adsabs.harvard.edu/abs/2006AJ....131.1163S} {131, 1163}

\bibitem[\protect\citeauthoryear{{Smartt}}{{Smartt}}{2009}]{Smartt09}
{Smartt} S.~J.,  2009, \mn@doi [\araa] {10.1146/annurev-astro-082708-101737}, \href {https://ui.adsabs.harvard.edu/abs/2009ARA&A..47...63S} {47, 63}

\bibitem[\protect\citeauthoryear{{Smartt}}{{Smartt}}{2015}]{Smartt15}
{Smartt} S.~J.,  2015, \mn@doi [\pasa] {10.1017/pasa.2015.17}, \href {https://ui.adsabs.harvard.edu/abs/2015PASA...32...16S} {32, e016}

\bibitem[\protect\citeauthoryear{{Smith}, {Pearson}, {Sand}, {Ilyin}, {Bostroem}, {Hosseinzadeh}  \& {Shrestha}}{{Smith} et~al.}{2023}]{Smith23}
{Smith} N.,  {Pearson} J.,  {Sand} D.~J.,  {Ilyin} I.,  {Bostroem} K.~A.,  {Hosseinzadeh} G.,   {Shrestha} M.,  2023, \mn@doi [arXiv e-prints] {10.48550/arXiv.2306.07964}, \href {https://ui.adsabs.harvard.edu/abs/2023arXiv230607964S} {p. arXiv:2306.07964}

\bibitem[\protect\citeauthoryear{{Soraisam} et~al.,}{{Soraisam} et~al.}{2018}]{Soraisam18}
{Soraisam} M.~D.,  et~al., 2018, \mn@doi [\apj] {10.3847/1538-4357/aabc59}, \href {https://ui.adsabs.harvard.edu/abs/2018ApJ...859...73S} {859, 73}

\bibitem[\protect\citeauthoryear{{Soraisam} et~al.,}{{Soraisam} et~al.}{2023}]{Soraisam23}
{Soraisam} M.~D.,  et~al., 2023, \mn@doi [arXiv e-prints] {10.48550/arXiv.2306.10783}, \href {https://ui.adsabs.harvard.edu/abs/2023arXiv230610783S} {p. arXiv:2306.10783}

\bibitem[\protect\citeauthoryear{{Soumagnac} et~al.,}{{Soumagnac} et~al.}{2020}]{Soumagnac20}
{Soumagnac} M.~T.,  et~al., 2020, \mn@doi [\apj] {10.3847/1538-4357/abb247}, \href {https://ui.adsabs.harvard.edu/abs/2020ApJ...902....6S} {902, 6}

\bibitem[\protect\citeauthoryear{{Speagle}}{{Speagle}}{2020}]{Speagle20}
{Speagle} J.~S.,  2020, \mn@doi [\mnras] {10.1093/mnras/staa278}, \href {https://ui.adsabs.harvard.edu/abs/2020MNRAS.493.3132S} {493, 3132}

\bibitem[\protect\citeauthoryear{{Srinivasan}, {Sargent}  \& {Meixner}}{{Srinivasan} et~al.}{2011}]{Srinivasan11}
{Srinivasan} S.,  {Sargent} B.~A.,   {Meixner} M.,  2011, \mn@doi [\aap] {10.1051/0004-6361/201117033}, \href {https://ui.adsabs.harvard.edu/abs/2011A&A...532A..54S} {532, A54}

\bibitem[\protect\citeauthoryear{{Teja} et~al.,}{{Teja} et~al.}{2023}]{Teja23}
{Teja} R.~S.,  et~al., 2023, \mn@doi [arXiv e-prints] {10.48550/arXiv.2306.10284}, \href {https://ui.adsabs.harvard.edu/abs/2023arXiv230610284S} {p. arXiv:2306.10284}

\bibitem[\protect\citeauthoryear{{Van Dyk}}{{Van Dyk}}{2017}]{VanDyk17}
{Van Dyk} S.~D.,  2017, \mn@doi [Philosophical Transactions of the Royal Society of London Series A] {10.1098/rsta.2016.0277}, \href {https://ui.adsabs.harvard.edu/abs/2017RSPTA.37560277V} {375, 20160277}

\bibitem[\protect\citeauthoryear{{Van Dyk} et~al.,}{{Van Dyk} et~al.}{2023a}]{VanDyk23_23ixf}
{Van Dyk} S.~D.,  et~al., 2023a, \mn@doi [arXiv e-prints] {10.48550/arXiv.2308.14844}, \href {https://ui.adsabs.harvard.edu/abs/2023arXiv230814844V} {p. arXiv:2308.14844}

\bibitem[\protect\citeauthoryear{{Van Dyk} et~al.,}{{Van Dyk} et~al.}{2023b}]{VanDyk23}
{Van Dyk} S.~D.,  et~al., 2023b, \mn@doi [\mnras] {10.1093/mnras/stac3549}, \href {https://ui.adsabs.harvard.edu/abs/2023MNRAS.519..471V} {519, 471}

\bibitem[\protect\citeauthoryear{{Vasylyev} et~al.,}{{Vasylyev} et~al.}{2023}]{Vasylyev23}
{Vasylyev} S.~S.,  et~al., 2023, \mn@doi [arXiv e-prints] {10.48550/arXiv.2307.01268}, \href {https://ui.adsabs.harvard.edu/abs/2023arXiv230701268V} {p. arXiv:2307.01268}

\bibitem[\protect\citeauthoryear{{Villaume}, {Conroy}  \& {Johnson}}{{Villaume} et~al.}{2015}]{Villaume15}
{Villaume} A.,  {Conroy} C.,   {Johnson} B.~D.,  2015, \mn@doi [\apj] {10.1088/0004-637X/806/1/82}, \href {https://ui.adsabs.harvard.edu/abs/2015ApJ...806...82V} {806, 82}

\bibitem[\protect\citeauthoryear{{Walmswell} \& {Eldridge}}{{Walmswell} \& {Eldridge}}{2012}]{Walmswell12}
{Walmswell} J.~J.,  {Eldridge} J.~J.,  2012, \mn@doi [\mnras] {10.1111/j.1365-2966.2011.19860.x}, \href {https://ui.adsabs.harvard.edu/abs/2012MNRAS.419.2054W} {419, 2054}

\bibitem[\protect\citeauthoryear{{Yamanaka}, {Fujii}  \& {Nagayama}}{{Yamanaka} et~al.}{2023}]{Yamanaka23}
{Yamanaka} M.,  {Fujii} M.,   {Nagayama} T.,  2023, \mn@doi [arXiv e-prints] {10.48550/arXiv.2306.00263}, \href {https://ui.adsabs.harvard.edu/abs/2023arXiv230600263Y} {p. arXiv:2306.00263}

\bibitem[\protect\citeauthoryear{{Yaron}, {Bruch}, {Chen}, {Irani}, {Zimmerman}, {Gal-Yam}  \& {Qin}}{{Yaron} et~al.}{2023}]{Yaron23TNS}
{Yaron} O.,  {Bruch} R.,  {Chen} P.,  {Irani} I.,  {Zimmerman} E.,  {Gal-Yam} A.,   {Qin} Y.,  2023, Transient Name Server AstroNote, \href {https://ui.adsabs.harvard.edu/abs/2023TNSAN.133....1Y} {133, 1}

\bibitem[\protect\citeauthoryear{{Yoon} \& {Cantiello}}{{Yoon} \& {Cantiello}}{2010}]{Yoon10}
{Yoon} S.-C.,  {Cantiello} M.,  2010, \mn@doi [\apjl] {10.1088/2041-8205/717/1/L62}, \href {https://ui.adsabs.harvard.edu/abs/2010ApJ...717L..62Y} {717, L62}

\bibitem[\protect\citeauthoryear{{Zimmerman}, {Irani}, {Chen}  \& {Gal-Yam}}{{Zimmerman} et~al.}{2023}]{Zimmerman23}
{Zimmerman} E.,  {Irani} I.,  {Chen} P.,   {Gal-Yam} A.,  2023, in prep.

\bibitem[\protect\citeauthoryear{{van Loon}, {Cioni}, {Zijlstra}  \& {Loup}}{{van Loon} et~al.}{2005}]{vanLoon05}
{van Loon} J.~T.,  {Cioni} M. R.~L.,  {Zijlstra} A.~A.,   {Loup} C.,  2005, \mn@doi [\aap] {10.1051/0004-6361:20042555}, \href {https://ui.adsabs.harvard.edu/abs/2005A&A...438..273V} {438, 273}

\makeatother
\end{thebibliography}








\bsp	
\label{lastpage}
\end{document}